\documentclass[12pt]{iopart}
\usepackage{graphicx}
\usepackage{iopams}
\newcommand{\bra}[1]{\langle{#1}|}
\newcommand{\ket}[1]{|{#1}\rangle}

\begin{document}

\title{Does Luttinger liquid behaviour survive in an atomic wire on a
  surface?}
\author{L K Dash and A J Fisher}
\address{Department of Physics and Astronomy, University College
  London, Gower Street, London WC1E 6BT}
\ead{\mailto{louise.dash@ucl.ac.uk}, \mailto{andrew.fisher@ucl.ac.uk}} 

\begin{abstract}
  We form a highly simplified model of an atomic wire on a surface by
  the coupling of two one-dimensional chains, one with
  electron-electron interactions to represent the wire and and one
  with no electron-electron interactions to represent the surface.  We
  use exact diagonalization techniques to calculate the eigenstates
  and response functions of our model, in order to determine both the
  nature of the coupling and to what extent the coupling affects the
  Luttinger liquid properties we would expect in a purely
  one-dimensional system.  We find that while there are indeed
  Luttinger liquid indicators present, some residual Fermi liquid
  characteristics remain.
\end{abstract}

\submitto{\JPCM}

\maketitle

\section{Introduction}
\label{sec:Introduction}
A purely one-dimensional metal is not expected to retain the
Fermi liquid properties of its three-dimensional counterpart, which is
characterized by a one-to-one relationship of its excitations to those
of a non-interacting Fermi system and an electron quasiparticle
lifetime that diverges to infinity near the Fermi surface.  Instead,
any interaction between electrons in a one-dimensional metal is
expected to lead to a non-zero scattering phase-shift and to destroy
completely the quasi-electron nature of the excitations \cite{Voit:95}.  The
resulting ``Luttinger liquid'' state has separate charge and spin excitations, and
quite different transport properties from those of a Fermi liquid.
The Luttinger liquid state is an adiabatic continuation of the ground state of a
simple one-dimensional model, the Luttinger model
\cite{Haldane:81-II}, in much the
same way that a Fermi-liquid state can be adiabatically derived from
the ground state of noninteracting fermions.

How could we observe this fascinating state of matter?  We cannot
fabricate purely one-dimensional systems in the laboratory, but we can
produce systems that are quasi-one-dimensional, to a greater or a
lesser degree.  In several of these, searches for Luttinger liquid behaviour have
been undertaken.  These include charge-transfer salts such as TMTSF
having almost one-dimensional bandstructures \cite{Schwartz:98}, semiconductor
quantum wires \cite{Wang:00}, edge states in the quantum Hall effect (which
are chiral systems along which electrons can only propagate in one
direction) \cite{Wen:90,Chang:96}, carbon nanotubes \cite{Bockrath:99}, and atomic-scale quantum
wires fabricated by self-assembly \cite{Segovia:99-I} or novel
SPM-based surface lithographies
\cite{Watanabe:97,Yajima:99,Hitosugi:99}.  The important question for
a theorist is whether such a quasi-one-dimensional system,
interacting with a three-dimensional environment, would be expected to
show Luttinger liquid behaviour or not, and if so, for what range of parameters.
This question was given particular timeliness by the recent
angle-resolved photoemission results of Segovia \etal \cite{Segovia:99-I}, which have
shown that photoemission from chains of Au atoms on Si$(111)$ exhibits
features which may be signatures of the spin and charge collective modes
indicative of Luttinger liquid behaviour.

In this paper we begin to construct a theoretical framework within
which we can answer these questions.  We consider the simplest
possible model of such a system --- one in which the atomic wire is
represented by a single-chain spinless Luttinger model, with
electron-electron interactions, on a lattice.  The environment is
represented only by a chain with no interactions but which is
otherwise identical.  Electrons can hop between the two chains by
hybridization of the electron states.  Such a simple system does not
capture the full complexities of electron scattering with the
environment, nor can it represent the magnetic behaviour of a
Luttinger liquid.  However it provides a relatively simple starting
point, and can already be used to calculate one-particle spectral
functions such as those probed by photoemission experiments.

In the following sections we explain the details of our model and our
numerical methods in Section~\ref{sec:Theoretical_methods}, and then
describe our results in Section~\ref{sec:Results}.  We find that some
aspects of the model display Luttinger liquid-like behaviour while others do not;
we then summarize these different properties in
Section~\ref{sec:Conclusions}.

\section{Theoretical methods}
\label{sec:Theoretical_methods}

\subsection{The Luttinger model}
\label{sec:luttinger-model}

\begin{figure}[htbp]
  \begin{center}
    \includegraphics[width=11cm]{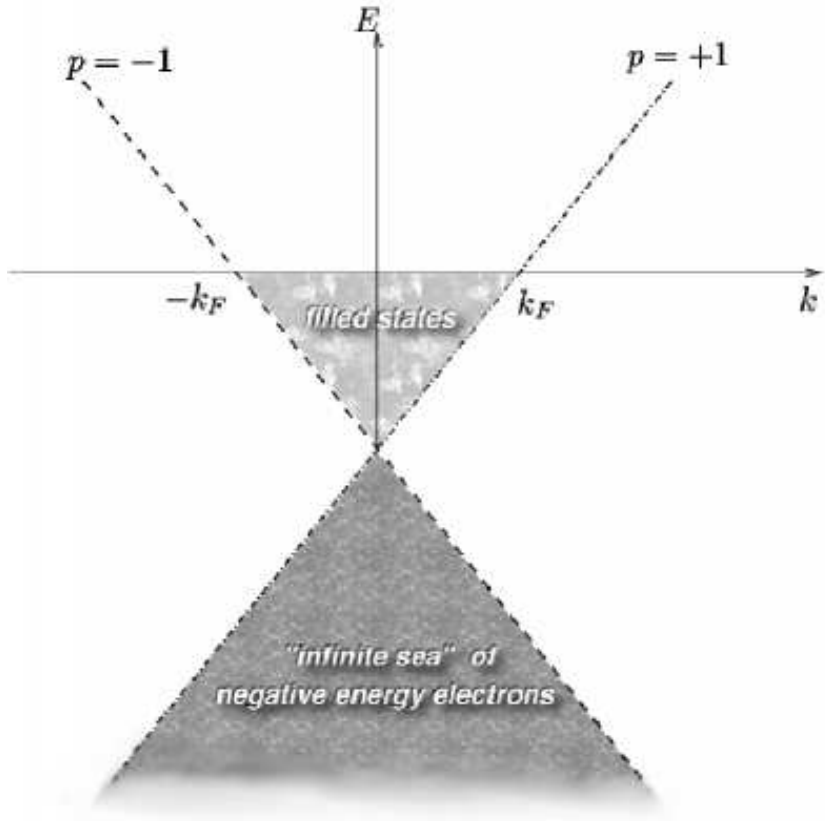}
    \caption{Schematic representation of the Luttinger model.  $p=+1,
      -1$ indicate the populations of left- and right-moving fermions
      respectively.}
    \label{fig:Luttinger_model}
  \end{center}
\end{figure}

We base our numerical calculations on Haldane's Luttinger model
operator method \cite{Haldane:81-II} for a chain of spinless idealized
atoms with periodic boundary conditions.  The Luttinger model is
characterized by a linearized dispersion relation with slope $v_{\rm F}$ and
a complete separation of the populations of left- and right-moving
particles (as in figure \ref{fig:Luttinger_model}) plus the inclusion
of an infinity of negative energy electrons
\cite{Haldane:81-II,Voit:95}. The second of these characteristics
makes the model exactly solvable by Haldane's bosonization method, as
the boson commutation relations are then exact rather than
approximate.
The total Luttinger model Hamiltonian can be split into three parts
\cite{Voit:95}:
\begin{equation}
  \hat{H} = \hat{H}_0 + \hat{H}_2 + \hat{H}_4. 
\end{equation}$\hat{H}_0$ represents the non-interacting part of the
Hamiltonian:
\begin{equation}
   \hat{H}_0 = \sum_{p,k} v_{\rm F}(pk-k_{\rm F})
   :\hat{c}^\dagger_{pk}\hat{c}_{pk}:,
\end{equation}
where $q$ indicates momentum and $p$ ($=\pm 1$) the branch
index for the left- ($-1$) and right-moving ($+1$) fermions.
The operator $\hat{c}^\dagger_{pk}$ creates a fermion of type $p$ with
momentum $k$ and $:\ldots:$ implies normal ordering.
The interaction parts of the Hamiltonian, $\hat{H}_2$ and
$\hat{H}_4$, are given respectively by
\begin{eqnarray}
  \hat{H}_2 =
   \frac{2}{L}\sum_{q}g_{2}(q)\hat{\rho}_{+}(q)\hat{\rho}_{-}(-q),
   \\
    \hat{H}_4 =
   \frac{1}{L}\sum_{p,q}g_{4}(q)\hat{\rho}_{p}(q)\hat{\rho}_{p}(-q).
\end{eqnarray}
The $\hat{H}_2$ term represents forward scattering between the left-
and right-moving fermion branches, while the $\hat{H}_4$ term
represents forward scattering within a momentum branch.  Note however
that $[\hat{H}_2,\hat{H}_0] \neq 0$, so $\hat{H}_2$ can modify the
ground state by excitation of particle-hole pairs, whereas $\hat{H}_4$
commutes with $\hat{H}_0$ and so cannot modify the ground state.  The
density operators $\hat{\rho}$ are defined in terms of the fermion
operators as
\begin{equation}
  \label{eq:rho}
  \hat{\rho}_{p}(q) = \sum_k
  :\hat{c}^\dagger_{p,k+q}\hat{c}_{p,k}: =
  \sum_{k}\left(\hat{c}^\dagger_{p,k+q}\hat{c}_{p,k} -
  \delta_{q,0}\langle\hat{c}^\dagger_{p,k}\hat{c}_{p,k}\rangle_0 \right).
\end{equation}

The complete Hamiltonian can be diagonalized via a Bogoliubov transformation
\cite{Haldane:81-II} to yield the bosonized form of the Hamiltonian
\begin{eqnarray}
   \label{eq:bosonized_hamiltonian}
\fl  \tilde{H} = \frac{\pi}{L}\sum_{p,\ q\neq 0} v_0(q)
  :\tilde{\rho}_p(q)\tilde{\rho}_p(-q): +
  \frac{\pi}{2L}\left[v_N\left(N_+ + N_-\right)^2 + v_J\left(N_+ -
      N_-\right)^2\right] \\
  \lo = \frac{1}{2}\sum_q\left(\omega_q -v_{\rm F}|q|\right)+ \sum_q\omega_q\hat{b}_q^\dagger\hat{b}_q +
    \frac{\pi}{2L}\left(v_NN^2 + v_JJ^2\right),
\end{eqnarray}
where the transformed density operators $\tilde{\rho}$ are related to the originals
by a phase $\phi_q$:
\begin{equation}
  \tilde{\rho}_p(q) = \hat{\rho}_p(q)\cosh \phi_q +
  \hat{\rho}_{-p}(q)\sinh\phi_q.
\end{equation}
There is a characteristic frequency $\omega_q = |(v_{\rm F} +
g_4(q)/2\pi)^2 - (g_2(q)/2\pi)^2|^{1/2}/|q|$ associated with the
transformed bosons, while the quantum numbers $N \equiv N_+ + N_-$ and
$J \equiv N_+ - N_-$ represent respectively the sum of the number of electrons on
the positive and negative branches, and the difference between them
(analogous to current).

The three velocities in the Hamiltonian are related as follows:
\begin{equation}\label{eq:velocities_1}
  \eqalign{
  v_N v_J = v_0^2, \\
  v_N = \frac{v_0}{K_\rho} = v_0 \rme^{-2\phi}, \\
  v_J = v_0 K_\rho = v_0 \rme^{+2\phi},
  }
\end{equation}
and are also related to the non-interacting Fermi velocity by
\begin{equation}\label{eq:velocities_2}
  \eqalign{
    v_N = v_{\rm F} + \frac{g_4 + g_2}{2\pi}, \\
    v_J = v_{\rm F} + \frac{g_4 - g_2}{2\pi}.
    }
\end{equation}
The requirement that $\tilde{H}$ be a diagonalized version of
$\hat{H}$ is ensured by the relationship between the Bogoliubov
transformation phase $\phi_q$ and the interaction functions $g_i$:
\begin{eqnarray}
  \label{eq:K_rho}
  \rme^{2\phi_q} &  = \left(\frac{\pi v_{\rm F} + g_4(q) -
  g_2(q)}{\pi v_{\rm F} + g_4(q) + g_2(q)}\right)^{1/2} \\
 & \equiv K_\rho(q).
\end{eqnarray}
The spinless Luttinger liquid parameter $K_\rho$ is then obtained by
taking the limit of $K_\rho(q)$ as $q$ tends to zero.  In all of the
above the limit $q\rightarrow 0$ is implied where $q$ is not
explicitly included.

In addition there are two further parameters related to $K_\rho$: for
a spinless Luttinger liquid they are given by
\begin{eqnarray}
  \alpha = \frac{1}{2}\left[K_\rho +
      \frac{1}{K_\rho} -2\right], \label{eq:alpha}\\
    \gamma = 2\alpha. \label{eq:gamma}
\end{eqnarray}
The Fermi liquid corresponds to $K_\rho = 1$ and $\alpha = 0$ and so
departures from these values can be used to ``measure'' the extent of
non-Fermi liquid behaviour. $\alpha$ is an exponent which governs the
power-law dependence of all single-particle properties (for example
the density of states, which varies as $N(E) \approx
|E-E_{\rm F}|^\alpha$), as well as other properties of the system,
including the d.c. resistivity, which varies as $\rho(T) \approx
T^{1-\gamma}$ \cite{Voit:95,Ogata:92}.

\subsection{Coupling two chains}
\label{sec:Coupling_two_chains}
We are now able to couple two Luttinger chains (labelled by superscript $A$ and
$B$) together by allowing hopping between adjacent points on
each chain with matrix element $t_\perp$:
\begin{equation}\label{eq:coupled_hamiltonian}
   \hat{H}^{\rm coupled} = \hat{H}^A + \hat{H}^B +
    t_\perp\sum_{x,p}\left(\hat{\psi}_p^{A\dagger}(x)\hat{\psi}_{p}^{B}(x) +
      \hat{\psi}_{p}^{B\dagger}(x)\hat{\psi}_{p}^{A}(x)\right).
\end{equation}
Other schemes of inter-chain hopping are of course possible, and we
expect that they will generate similar results.  We neglect any ``drag
effects'' of inter-chain interactions that would be generated by
interaction terms analogous to $\hat{H}_2$ or $\hat{H}_4$ involving
electrons on both chains.  We have the choice of using either a set of
basis states generated by the diagonalized boson operators
$\hat{b}_q^\dagger$ or the non-diagonalized operators
$\hat{a}_q^\dagger$.  We choose the latter, as although this has the
consequence that the ground states for a single interacting chain no
longer consist of the relevant zero-boson basis states, it is
considerably more convenient computationally.  $\hat{\psi}^\dagger$
are fermion creation operators given in the bosonized form by
\begin{equation}
  \label{eq:fermion_boson}
  \hat{\psi}_{p,c}^{\dagger}(x) = \frac{1}{\sqrt{L}} \rme^{\rmi pk_{\rm F}x}
  \left[\rme^{\rmi\hat{\phi}_p^{\dagger}(x)} \hat{U}_{p,c} \rme^{\rmi\hat{\phi}_p(x)}\right],
\end{equation}
$\hat{\phi}_p(x)$ is a boson field operator given by
\begin{equation}
  \label{eq:boson_field_operator}
  \hat{\phi}_p(x) = \left(\frac{p \pi x}{L}\right)N_p + 
  i\sum_{q\neq 0}\theta(pq)\left(\frac{2\pi}{L|q|}
  \right)^{1/2}\rme^{\rmi qx}\hat{a}_q,
\end{equation}
where the subscripts $p$ refers to the branch index ($\pm 1$) and $c$
to the chain index.  $\hat{U}_{p,c}$ is a ladder operator whose form
ensures the anticommutation properties of the final fermion operator $
\hat{\psi}_p^{\dagger}(x)$ despite the commutation properties of the
constituent boson operators $\hat{a}_q$.

In order for $\hat{U}_{p,c}$ to produce anticommuting field operators
on different chains, it is necessary to introduce a further phase
factor into its definition, analogous to Haldane's phase factor
$\zeta(p,N_p,N_{-p})$ for ensuring anticommutation between the
branches of a single chain.  The total ladder operator component of
equation (\ref{eq:fermion_boson}) thus takes the form
\begin{equation}\label{eq:ladderoperator}
  \hat{U}_{p,c} =
  \zeta(p,N_p,N_{-p})\zeta^\prime(c,N_c,N_{-c})|N_{p,c}+1, N_{-p,c}\rangle,
\end{equation}
where the subscript $c = \pm 1$ is a chain index. The anticommutation
properties originate in the phase factors $\zeta$, which can be
written as
\begin{equation}\label{phase_factor_zeta}
  \zeta_{i=p,c} = (-1)^{(\frac{1}{2}iN_{-i})}.
\end{equation}

\subsection{Computational details}
\label{sec:Computational_details}

The choice of a specific form for the interactions $g_2(q)$ and
$g_4(q)$ is arbitrary, as
the Luttinger liquid remains completely solvable for any interaction
that fulfils certain conditions \cite{Haldane:81-II}.  We choose a
Gaussian form for the interaction and set $g_2(q) = g_4(q) = 2\pi V(q)$ with
\begin{equation}
  \label{eq:gaussian_interaction}
  V(q) = I \exp{\left(-2q^2/r\right)}
\end{equation}
as this has the advantage that it
maintains the same form in both real and momentum space. Other forms,
such as a screened Coulomb interaction, would also be possible.  The parameter
$I$ can be varied to control the ``strength'' of the
interaction, and $r$ controls the range of the interactions.

Our model is completed by setting the interaction strength on chain B
to zero but retaining interactions on chain A. With the
non-interacting chain representing the surface and the interacting
chain the wire, we therefore form our highly simplified model of an
atomic wire interacting with a substrate.
\begin{figure}
  \begin{center}
    \includegraphics[width=11cm]{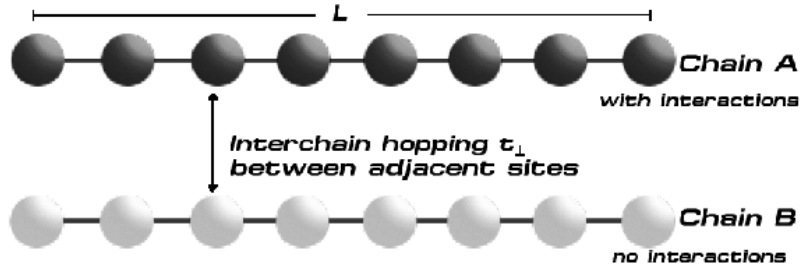}
    \caption{Representation of the two chains.  Electron-electron
      interactions are included on just one of the chains.}
    \label{fig:chain_hopping}
  \end{center}
\end{figure}

The size of the computational Hilbert space is restricted by allowing
only one boson in each mode.  While this successfully truncates the
Hilbert space to a manageable size it does not affect the low energy
properties in which we are interested, as we have checked that only
higher energy excitations involve basis states with more than one
boson in each mode.  We also neglect the presence of the infinity of
negative energy electrons, i.e.
$\hat{c}\ket{N_\textrm{\scriptsize{electrons}}=0} = 0 $.  However, we
are still strongly limited in the size of the system we can handle:
the Hilbert space scales exponentially with both length of chains and
number of electrons.  The largest system for which we have been able
to calculate with arbitrary numbers of electrons is one with length $L
= 6$ (i.e. 12 sites in total).  In order to obtain both the
eigenstates of the coupled system and its correlation functions we
utilize the Lanczos method \cite{Haydock:80,GolubVanLoan:96}.

\subsection{Spectral functions}
\label{sec:Spectral_functions_theory}
We define the spectral function $\rho(q,\omega)$ as 
\begin{equation}
  \label{eq:spectral_function}
  \rho(q,\omega)_{p,c} = -\frac{1}{\pi} \Im\left[ G^R_{p,c}(k_{\rm
 F} + q, \omega + \mu)\right],
\end{equation}
with $q=k_{\rm F}-k$, and the retarded Green's function $G^R_{p,c}$ defined
as the double Fourier transform of
\begin{equation}
  \label{eq:retarded_greens_function}
   G^R_{p,c}(x,t) = -\rmi\theta(t)\{ \langle \hat{\psi}_{p,c}(x,t)
   \hat{\psi}_{p,c}^\dagger(0,0)\rangle \}
\end{equation}
where the subscripts $p$ and $c$ again indicate branch and chain
indices.  It is possible to derive an analytic form for the spectral
function of a single Luttinger liquid, either with or without spin, as
in references \cite{Schoenhammer:93,Voit:93,MedenSchonhammer:92}.
Whereas for a Fermi liquid we would expect $\rho(0,\omega)$ to be a
delta function at the Fermi energy, the situation for a Luttinger
liquid is quite different.  Spectral weight is repelled from the Fermi
surface owing to the virtual particle-hole excitations generated by
the inter-branch scattering term $g_2$, resulting in a broadened peak.
As $q$ is increased the Fermi liquid spectral function will merely
broaden like $q^2$ reflecting the finite lifetime of electrons away
from $E_F$, but for a Luttinger liquid there is zero spectral weight
within a range $\pm v_0 q$ of the Fermi energy.  In addition, the
negative frequency contribution is suppressed exponentially with $q$,
and for a continuum Luttinger liquid the positive frequency
contributions have a power law dependence.  For $\omega>0$ this is of the form
$\theta(\omega-v_0q)(\omega-v_0q)^{\gamma-1}$ and for $\omega<0$ it is
of the form $\theta(-\omega-v_0q)(-\omega-v_0q)^\gamma$, with $\gamma$
given by equation (\ref{eq:gamma}) \cite{Voit:95}.

However, for our more complex system it is not easy to obtain an
analytic form and so we resort to computational methods.  We calculate
the Green's function
\begin{equation}
  \label{eq:Green's_function_Haydock}
\fl  G^R(k,k^\prime,\omega) = \bra{N}\hat{c}_k \left[\omega - \hat{H} +
  \varepsilon_N \right]^{-1}\hat{c}^\dagger_{k^\prime}\ket{N} +
  \bra{N} \hat{c}^\dagger_{k^\prime}\ \left[\omega + \hat{H} -
  \varepsilon_N \right]^{-1}\hat{c}_k\ket{N}
\end{equation}
using Haydock's tridiagonal Lanczos-based procedure
\cite{Haydock:80}.  In order to ensure convergence, we put $\omega
\rightarrow \omega +\rmi\eta$, where $\eta$ is an imaginary component
of the energy roughly equal to the level spacing of the system
\cite{Haydock:80}. $\ket{N}$ is the $N$-electron ground state, with
energy $\varepsilon_N$.

\section{Results}
\label{sec:Results}
In this section we present our numerical results, concentrating on
$L=6$ (our largest system).  The results for $L=4$ are very similar,
and this encourages us to believe that, despite the small size of our
system, finite-size effects are not dominating the results.  All
results shown here were calculated with $v_{\rm F}=1$. 

\subsection{Ground and excited states}
\begin{figure}[h]
  \begin{center}
  \includegraphics[width=11cm]{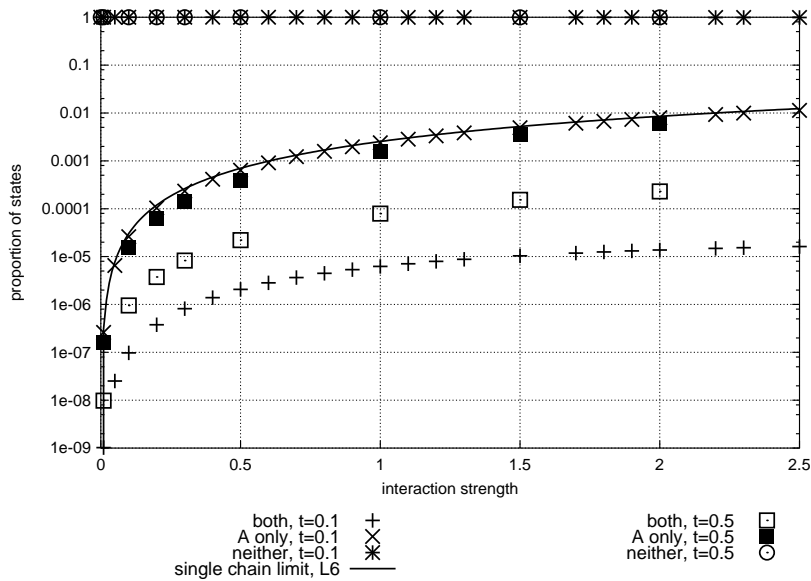}
  \caption{Contribution of bosonic basis states to the ground states
    of the $L=6$, $N=2$, $t_\perp=0.1$ and $L=6$, $N=2$, $t_\perp=0.5$
    systems as a function of
    interaction strength $I$.  See text for a full explanation.}
  \label{fig:bosoncontributions}
  \end{center}
\end{figure}

The ground and neutral first excited states are calculated for
different values of the interaction strength $I$ as detailed above.
As we have chosen the basis states for the system to be those before
the Bogoliubov transformation, we find that the non-interacting ground
state (i.e. $I=0$) consists entirely of zero-boson basis states,
whereas all the interacting systems ($I>0$) have contributions from
basis states that include bosons on one or both chains, as these basis
states are not individually eigenstates of the interacting system.
The contribution to the ground states from basis states containing
either no bosons on either chain, bosons on the chain with explicit
interactions only (i.e. chain A) and bosons on both chains are shown
in figure \ref{fig:bosoncontributions}.

\label{sec:Ground_and_excited_states}

The corresponding data for the boson contributions to a single
isolated Luttinger liquid chain with otherwise identical parameters is also shown in
figure \ref{fig:bosoncontributions}.  This matches almost exactly the
contribution for the system's interacting chain (chain A) and thus
indicates we are in a regime where the the two chains can be said to
be weakly coupled for both coupling strengths $t_\perp = 0.1$ and
$0.5$.
\begin{figure}
  \begin{center}
  \includegraphics[width=11cm]{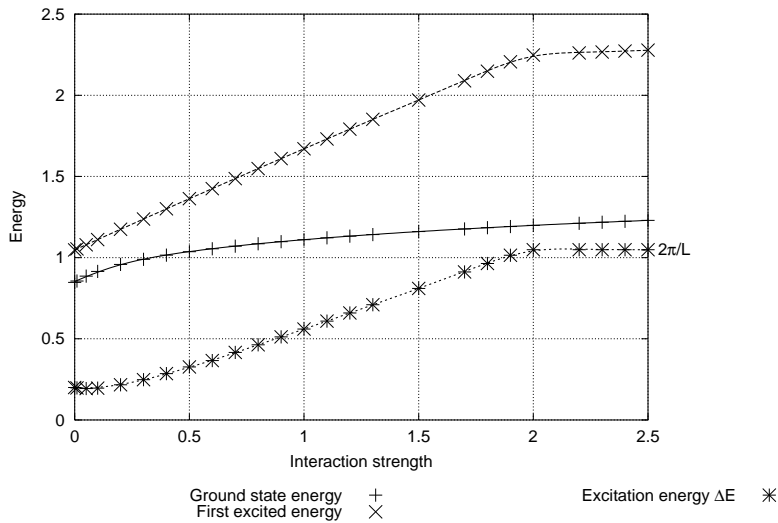}
  \caption{Energies for the ground state and neutral first excited state
    for the $L=6$, $N=2$, $t_\perp=0.1$ system as a function of
    interaction strength $I$.}
  \label{fig:energy_scaling}
  \end{center}
\end{figure}

The energies of the ground and first excited states are plotted in
figure \ref{fig:energy_scaling}.  We see that the ground state energy
has a region of roughly linear dependence on $I$ in the range
$0\lesssim I \lesssim 0.5$ and thereafter is rather less dependent
on $I$.  The first excited state, however, is linear in $I$ for
$I\lesssim 2$, at which point it becomes completely independent of
$I$.  The excitation energy $\Delta E$ for the non-interacting system
is equal to $2t_\perp$, indicating that for this size of system the
first excitation is one between the bonding and anti-bonding bands
with zero net momentum change.  As we increase $I$ however, the
bonding and antibonding bands distort to such an extent that
eventually, at $I \approx 2.0$, the lowest excitation becomes not one
between bands with zero net momentum, but an intra-band one with net
momentum change $2\pi v/L$.  For a Luttinger liquid, we would
expect the velocity $v$ in this expression to be the Luttinger
velocity $v_0$ (as in equations
(\ref{eq:velocities_1}), etc.) 
rather than the Fermi velocity $v_{\rm F}$, leading to a continued
$I$-dependence for $\Delta E$.  However, it is clear from figure
\ref{fig:energy_scaling} that this is not the case for this system:
the excitation energy for $I\gtrsim 2.0$ remains constant and equal to
$2\pi v_{\rm F}/L = \pi/3$. (For this system $v_{\rm F} = 1$.)  This
can be explained as follows --- as $I$ is increased the single-chain
energy levels on chain A increase (since the interaction is
repulsive), while those on the non-interacting chain B remain
constant.  It therefore becomes energetically preferable for the
electrons to move to the non-interacting chain B rather than remain on
the (interacting) chain A, until the point is reached at which the
system is predominantly populated by electrons on chain B.  This chain
is still Fermi liquid-like, thus shifting the properties of the system
as a whole back from partially Luttinger liquid-like to Fermi
liquid-like.

\begin{figure}
  \begin{center}
  \includegraphics[width=11cm]{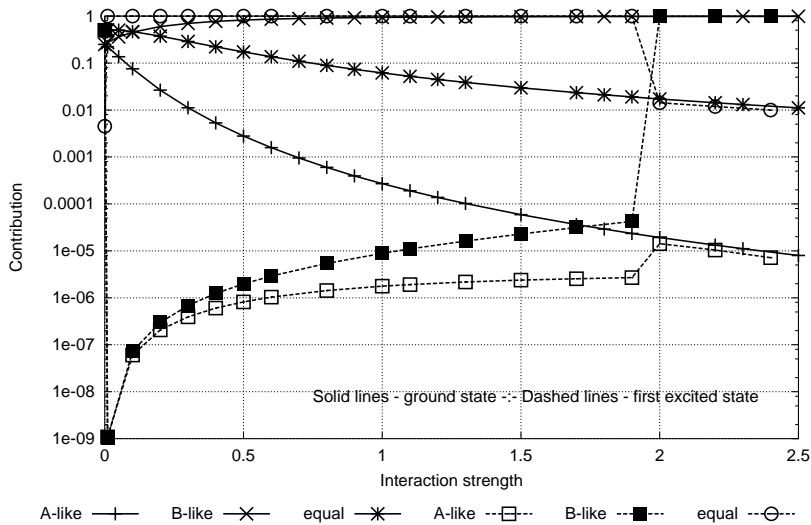}
  \caption{Contributions to the ground and first excited states for
    the $L=6$, $N=2$, $t_\perp=0.1$ system from electrons on each
    chain as a function of interaction strength $I$.  See text for a
    full explanation.}
  \label{fig:charge_distribution}
  \end{center}
\end{figure}

This is confirmed by an examination of the location of the electrons
for both the ground and first excited states.  Figure
\ref{fig:charge_distribution} shows a plot for $L=6, N=2,
t_\perp=0.1$,
showing the chain to which the electrons ``belong'', again based on an analysis
of the contribution of each of the basis states forming the
eigenstate.  As we have only 2 electrons in this system, there are
three possible distributions for the electrons: both on chain A
(``A-like''), both on chain B (``B-like'') or one on each chain
(``equal'').

First consider the ground states.  We see that for the non-interacting
ground state the electrons are distributed equally between the chains,
and once interactions are switched on we evolve continuously towards a
regime where basis states with electrons on chain A are strongly
suppressed in favour of states with both electrons on chain B.

The situation for the excited states is however quite different.  We
start at $I=0$ with a state in which, like the ground state, the
electrons are equally distributed between the chains.  This state is,
however, 4-fold degenerate and immediately even weak interactions are
introduced, this degeneracy is lifted and, in what is now the lowest
state, the contribution from basis states with both
electrons on the same chain are strongly suppressed.  The first
excited state for small values of $I$ may therefore be thought of as
``covalent'' in nature, as opposed to the ``ionic''-like states that
contribute to the non-interacting state where both electrons are on
the same chain.  As interactions are increased further, the
contribution from the ionic-like states begins to return.  However,
once we hit the point (at $I = 2.0$) where the nature of the
excitations changes from zero net momentum change to $\Delta
k= 2\pi v_{\rm F}/L$, there is once more a discontinuity in the plot
and the electron contributions from this point match those of the
ground state, indicating that there is no net interchain electron
hopping involved in these excitations.

\subsection{Spectral functions}
\label{sec:Spectral_functions_results}
\begin{figure}[htbp]
  \begin{center}
    \includegraphics[width=11cm]{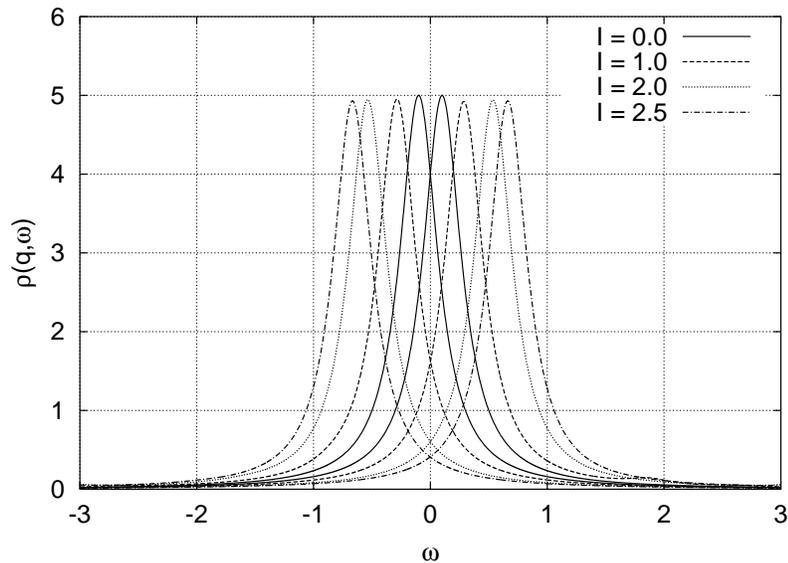}
    \caption{Hole (left peak) and electron (right peak) contributions
      to the spectral function $\rho(q=0,\omega)$.
      $t_\perp = 0.1$, $\eta = 2t_\perp$.  In order to show the splitting
      clearly, the electron and hole contributions are shown as equal
      in magnitude.}
    \label{fig:Spectral_function_q=0}
  \end{center}
\end{figure}

Results from the calculation of the spectral function
$\rho(q=0,\omega)$ are presented in figure
\ref{fig:Spectral_function_q=0}.  There are three sources of the
broadening of the peaks away from the delta function one would expect
from a Fermi liquid.  The first of these is due to the finite value of
the imaginary energy $\eta$ we have used in our calculations, leading
to broadening of the peaks into Lorentzians with width $\sim \eta$.
The second results from the splitting into bonding and anti-bonding
bands due to the inter-chain coupling and is equal to $2t_\perp$.  The
remaining broadening, however, is due to the removal of
spectral weight from the Fermi energy at $\omega=0$, and is indicative
of Luttinger liquid behaviour in the coupled system.

\begin{figure}[htbp]
  \begin{center}
    \includegraphics[width=11cm]{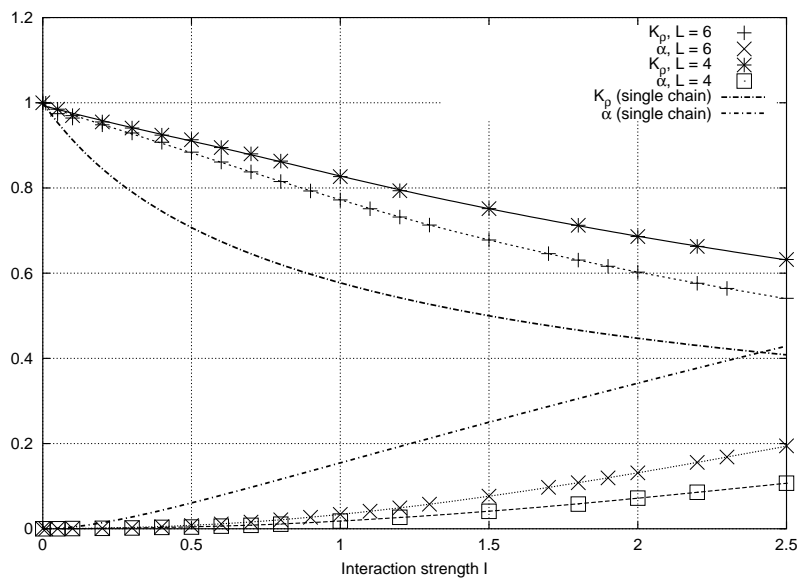}
    \caption{Luttinger parameters $K_\rho$ and $\alpha$ for the
      coupled chain system, both $L = 6$ and $L = 4$, and for an
      isolated continuum limit ($L\rightarrow \infty$) chain.}
    \label{fig:Luttinger_parameter_results}
  \end{center}
\end{figure}

We can also calculate $\rho(q\neq 0, \omega)$ although for our
extremely small system, we are severely restricted in available values
of $q$.  As we would expect for a Luttinger liquid, there is a strong
suppression of the hole contribution for $q>0$. Owing to the discrete
nature of this system, it is not possible to calculate the Luttinger
parameter $K_\rho$ direct from the slopes of the electron and hole
contributions as in references
\cite{Schoenhammer:93,Voit:93,MedenSchonhammer:92}.  However, we can
still derive a value for $K_\rho$ from the separation of the electron
and hole peaks as follows: For a non-interacting coupled chain system,
the electron and hole contributions to $\rho(q= 0, \omega)$ are
separated by $2t_\perp$, whereas for a single Luttinger liquid the
contributions to $\rho(q\neq 0, \omega)$ are separated by $2v_0 q$.
We define an effective $v_0$ in terms of the peak splitting
$\Delta\omega$ by the relation
\begin{equation}
  \label{eq:peak_splitting}
  \Delta\omega = 2(v_0^{\rm eff}q + t_\perp).
\end{equation}
Then, since we have $g_2(q) = g_4(q)$, we can use $v_0^{\rm eff}$ as a
measure of $v_0$ and equate equations (\ref{eq:velocities_1}) and
(\ref{eq:velocities_2}) to give an expression for the Luttinger
parameter:
\begin{eqnarray}
  \label{eq:K_rho_results}
  K_\rho^{-1}  & = \rme^{-2\phi} =
  \frac{1}{v_0^{\rm eff}} \left[v_{\rm F} + \frac{({v_0^{\rm eff}}^2 -
  v_{\rm F}^2)}{v_{\rm F}}\right] \\
   & = v_0^{\rm eff},
\end{eqnarray}
since $v_{\rm F}=1$.
The results of this calculation and the corresponding values of the
parameter $\alpha$ (equation (\ref{eq:alpha})) are plotted in figure
\ref{fig:Luttinger_parameter_results}.

\section{Conclusions}
\label{sec:Conclusions}
Our calculations approach the question of whether our model exhibits
Luttinger liquid behaviour from several complementary directions.  We
have presented results for the nature of the ground state, for the
nature of excited states (charged and neutral) and the related
response functions governing the response of the system to excitation.
We have also estimated the exponents in certain power-laws
characterizing the Luttinger liquid state.

Our results display a certain ambiguity with regard to whether or not
this system is a Luttinger liquid.  Roughly speaking, the evidence
from the bosonic nature of the ground state, the charged excited
states, and the values of the scaling parameters is all broadly
consistent with Luttinger liquid behaviour, modified by the presence of the second
chain.

Specifically, the one-particle spectral function is consistent with
Luttinger liquid behaviour.  At $q=0$, the total spectral function is
broadened by separation of the electron and hole parts---this
separation is much greater than the $2t_\perp$ expected in a
non-interacting two-chain system.  For $q\ne0$, there is additional
separation of the electron and hole parts, consistent with the further
suppression of spectral weight near the Fermi energy.  (For $q>0$,
corresponding to probing the occupancy of states with $k>k_{\rm F}$ --- see
equation~(\ref{eq:spectral_function}) --- the hole part of the spectral
function is strongly suppressed, even for relatively strong
interactions.)  Further support for the Luttinger liquid hypothesis comes from
the composition of that portion of the ground state in which bosons
are excited on chain A: this fraction closely tracks the fraction that
would be expected in an isolated single-chain Luttinger liquid 
(Figure~\ref{fig:bosoncontributions}).

Moreover, we are able to derive indirectly from $\rho(q>0,\omega)$
values for two Luttinger liquid parameters: $K_\rho$ (which determines the
ratio of the different velocities) and $\alpha$ (which determines the
asymptotic behaviour of the correlation functions).  Although our
analysis presupposes the validity of relations characterizing the Luttinger liquid
state (equation (\ref{eq:K_rho_results})), the results are
self-consistent in the sense that they show $K_\rho<1$ and $\alpha>0$,
i.e., departures from the Fermi-liquid values.  The degree of
departure from Fermi-liquid values is, however, somewhat less than
would be expected for a single-chain system with the same interaction
strength (see Figure~\ref{fig:Luttinger_parameter_results}).

However, we also have evidence that there is some residual Fermi
liquid behaviour.  This comes principally from the distribution of
charge across the two chains, as a function of interaction strength, in
the ground and neutral excited states.  Transfer of charge to the
non-interacting chain at higher values of $I$ results in the
resurrection of a Fermi-liquid like neutral excitation spectrum
(although we are still able to extract non-Fermi liquid values for the
Luttinger parameters for these interaction strengths, as explained
above).

This implies that while the neutral excitation spectrum
undergoes a transformation to Fermi liquid behaviour at large $I$, the
charged excitation spectrum does not.  It remains to be
seen how the apparent conflict between these different views on the
same underlying quantum state is resolved.

Finally, it should be noted that in our system the surface is
represented as metallic, although in order to achieve true
one-dimensional conduction in an experiment it is much more likely
that a semiconducting or insulating substrate would be employed, as in
the work of \cite{Segovia:99-I}.  One might expect that with a gapped
substrate, the elimination of hybridization with substrate states at
the Fermi energy would promote Luttinger liquid formation by comparison with the
metallic case.  So our finding that some aspects of Luttinger liquid spectral
behaviour persist even in the metallic case may also have implications
for recent and future experiments.  We plan further theoretical work
on semiconducting substrates.  It would also be interesting to study
the influence of wire-substrate interactions other than simple
hybridization on Luttinger liquid formation.

\ack{LKD thanks EPSRC and the National Physical Laboratory for support
  in the form of a CASE studentship;  AJF thanks EPSRC for the award
  of an Advanced Fellowship. \\}


\end{document}